\documentclass[aps,prd,showpacs,twocolumn,
superscriptaddress]{revtex4-1}

\usepackage{graphicx}
\usepackage{subfigure} 
\usepackage{dcolumn}

\graphicspath{{figures/}}
\usepackage{amsmath}
\usepackage{amssymb}
\usepackage{bm}
\usepackage{color}
\usepackage{comment}

\usepackage[normalem]{ulem}
\usepackage[dvipsnames]{xcolor}
\usepackage{hyperref}
\usepackage{multirow}

\providecommand{\mnras}{Mon. Not. R. Astron. Soc.}
\providecommand{\nat}{Nature}
\providecommand{\apj}{Astrophys. J.}
\providecommand{\apjl}{Astrophys. J. Lett.}
\providecommand{\prd}{Phys. Rev. D}
\providecommand{\prl}{Phys. Rev. Lett.}

\hypersetup{
colorlinks=true, 
linkcolor=blue,  
citecolor=cyan,  
}

\newcommand{\bea}{\begin{eqnarray}}
\newcommand{\eea}{\end{eqnarray}}
\newcommand{\be}{\begin{equation}}
\newcommand{\ee}{\end{equation}}
\graphicspath{{figures/}}

\begin{document}

\title{Measurability of Quadrupole Deviations from Kerr in Binary black hole Mergers}

\author{Song Li}
\email{leesong@shao.ac.cn}
\affiliation{Shanghai Astronomical Observatory, Shanghai, 200030, China }
\affiliation{School of Astronomy and Space Science, University of Chinese Academy of Sciences,
Beijing, 100049, China}

\author{Wen-Biao Han}
\email{Corresponding author: wbhan@shao.ac.cn}
\affiliation{Shanghai Astronomical Observatory, Shanghai, 200030, China }
\affiliation{School of Fundamental Physics and Mathematical Sciences, Hangzhou Institute for Advanced Study, UCAS, Hangzhou 310024, China }
\affiliation{School of Astronomy and Space Science, University of Chinese Academy of Sciences,
Beijing, 100049, China}
\affiliation{International Centre for Theoretical Physics Asia-Pacific, Beijing/Hangzhou, China}
\affiliation{Shanghai Frontiers Science Center for  Gravitational Wave Detection, 800 Dongchuan Road, Shanghai 200240, China}

\date{\today}

\begin{abstract}
We investigate the measurability of black hole quadrupole deviations from Kerr using five binary black hole mergers observed by the LIGO-Virgo-KAGRA Collaboration with the beyond-general-relativity full-waveform model $\Psi_{\mathrm{FD}}$. While earlier lower-SNR events mildly favored nonzero quadrupole deviations, the newly included high-SNR GWTC-4 events GW231226, GW230814, and GW250114 yield results increasingly consistent with the Kerr prediction. In particular, GW230814 and GW250114, the two highest-SNR events in our sample, yield deviations consistent with zero. We further perform separate inspiral and post-inspiral analyses and find both the posterior distributions centered close to $\Delta Q/Q=0$ for GW230814 and GW250114. Overall, the full-waveform, inspiral, and post-inspiral results for GW230814 and GW250114 reveal no observable departure from the no-hair theorem within the sensitivity of the current data and the $\Psi_{\mathrm{FD}}$ framework. Although the limited number of events prevents a definitive conclusion, future detections of additional high-SNR binary black hole mergers will enable increasingly stringent and robust tests of the Kerr nature of black holes.
\end{abstract}

\maketitle

\section{Introduction}
In November 1915, Albert Einstein finalized his formulation of gravitation with the publication of the theory now known as general relativity (GR), providing a revolutionary description of gravity as the manifestation of spacetime curvature. For several decades following its inception, GR remained largely a theoretical construct, admired for its mathematical elegance but supported by limited empirical validation. It was not until the technological advances of the 1960s that systematic experimental tests of gravity became feasible, marking the emergence of modern experimental gravitation\cite{Will_14}. Since then, an extensive array of observations has confirmed the predictions of GR with remarkable precision, encompassing laboratory experiments, Solar System tests, and observations of compact stellar systems\cite{Schlamminger_08, Wagner_12, Bertotti_03, Pitjeva_03, Williams_04, Kramer_06, Taylor_82, Damour_96}. Particularly compelling confirmations have arisen from high-resolution radio imaging in the vicinity of supermassive black holes\cite{EHT_1, EHT_2, EHT_3, EHT_4, EHT_5}, which probe gravity in some of the strongest accessible gravitational field regimes.

Among the most striking predictions of general relativity in the strong field regime is the black hole no-hair theorem. This theorem states that astrophysical black holes formed through gravitational collapse and settling into a stationary equilibrium are completely characterized by only two parameters: their mass and angular momentum, and are uniquely described by the Kerr metric. There are also some other non-Kerr metrics with naked singularities or other pathologies \cite{Ricci_1, Ricci_2, Ricci_3}, recently, researchers have studied Zipoy-Voorhees spacetime (also known as $\delta$-metric,$q$-metric or $\gamma$-metric) \cite{ZV_1, ZV_2, ZV_3, ZV_4, ZV_5}, while for a more general metric solution, they have proposed and studied the $\delta$-Kerr metric \cite{deltaKerr_1,deltaKerr_2,deltaKerr_3}. This nonlinear superposition of the Zipoy-Voorhees metric with the Kerr metric represents a deformed Kerr solution. The no-hair theorem, also the Kerr metric, therefore provides a sharp and falsifiable prediction of GR, making it a central target for observational tests of general relativity. Any robust evidence for deviations from the Kerr geometry would signal a violation of the no-hair theorem and potentially point to new physics beyond GR.

Over the past decade, significant advancements in gravitational-wave astronomy have made such tests feasible. The first detection of a binary black hole merger, GW150914, by the LIGO–Virgo Collaboration in 2015 opened the era of observational strong-field gravity\cite{GW150914}. Since then, the LIGO–Virgo-KAGRA Collaboration has unearthed more than 200 compact binary mergers until the first part of the fourth observing run (O4a)\cite{O1, O2, O3, O4_1, O4_2, O4_3}. Most of them are binary black hole (BBH) mergers, but some are binary neutron star (BNS)\cite{BNS} and neutron star black hole (NSBH) mergers\cite{NSBH}. These high signal-to-noise ratio events provide excellent opportunities to test GR and the no-hair theorem and to gain new insights into the nature of compact objects\cite{Test_GR_1, Test_GR_2, Test_GR_3, Test_GR_4, Test_GR_5, Test_GR_6, Test_GR_7}. Ground-based detectors such as LIGO, Virgo, and KAGRA have been crucial in advancing our understanding of compact objects, and upcoming space-based missions like the Laser Interferometer Space Antenna (LISA), Taiji\cite{taiji}, and Tianqin\cite{tianqin} will further enhance our ability to probe strong-field gravity across the Universe.

Various methods have been employed to test the black hole no-hair theorem using different observational facilities, including ground-based gravitational-wave detectors and the Event Horizon Telescope (EHT). Isi \cite{Test_GR_2} analyzed the gravitational-wave event GW150914, focusing on the ringdown portion of the signal, and reported a test of the no-hair theorem at the $\sim 10\%$ level. Abbott et al. \cite{Abbott_21} constrained the spin-induced quadrupole moment deviation parameter $\delta \kappa_s$, obtaining upper limits of 11.33 and 110.89 at the 90\% confidence level for GW151226 and GW190412, respectively. Broderick et al. \cite{Broderick_2014} employed a quasi-Kerr spacetime, which allows for an independent quadrupole moment, to investigate potential deviations from the Kerr metric through simulated images of Sgr A*. Although this work presented the first simulated images of a radiatively inefficient accretion flow (RIAF) around Sgr A* in a quasi-Kerr metric, the resulting constraints on departures from the Kerr geometry were relatively weak. More recently, the exceptionally high signal-to-noise ratio (SNR) event GW250114, with an SNR of approximately 80, provides a unique opportunity to perform precision tests of general relativity and Hawking’s area theorem \cite{HK_25}.


Most studies testing the no-hair theorem using gravitational waves rely on waveform templates. If observed gravitational-wave events are consistent with the predictions of general relativity, this provides support for the validity of the no-hair theorem. The gravitational waveform from a binary black hole coalescence can be divided into three stages: inspiral, merger, and ringdown. During the inspiral phase, the black holes gradually approach each other under the emission of gravitational radiation. In the merger phase, the two black holes coalesce to form a single remnant, while in the ringdown phase, the remnant undergoes damped oscillations, radiating excess energy and angular momentum until it reaches a stable equilibrium. Different theoretical approaches are employed to describe each stage. The inspiral phase is typically modeled using post-Newtonian theory, which accurately captures the slow orbital decay of the binary. The merger phase requires numerical relativity simulations to resolve the highly nonlinear dynamics of the coalescence. The ringdown phase, corresponding to small perturbations of the remnant black hole spacetime, is naturally described using quasinormal modes (QNMs), which characterize the damped oscillations and decay of the gravitational-wave signal. Well-established waveform templates, such as SEOBNR \cite{EOB_1, EOB_2, EOB_3} and IMRPhenom \cite{IMRPhenom_1, IMRPhenom_2, IMRPhenom_3}, combine these approaches to provide accurate inspiral–merger–ringdown predictions. If a waveform template can encode potential deviations from the Kerr geometry, it can be used to quantify and test violations of the no-hair theorem. In previous work \cite{Li_23}, we constructed a non–general-relativity full waveform template, denoted $\Psi_{\mathrm{FD}}$. In this paper, we employ this template to perform systematic tests of the no-hair theorem.

This article is organized as follows. In Sec.~\ref{Waveform_template_sec}, we introduce the waveform template $\Psi_{\mathrm{FD}}$. Sec.~\ref{PE_sec} describes the parameter estimation methodology. The results of our analysis are presented in Sec.~\ref{Results_sec}. Finally, Sec.~\ref{Conculsion_sec} summarizes the main findings and conclusions. Throughout this paper, we use geometrical units $G=c=1$.

\section{waveform template}\label{Waveform_template_sec}
The gravitational waveform template employed in this paper is the $\Psi_{\mathrm{FD}}$ model, a non–general-relativity full waveform template applicable to arbitrary axisymmetric black holes\cite{Li_23}. In the following, we briefly review the main features of the $\Psi_{\mathrm{FD}}$ model. Its public implementation is available at \footnote{\url{https://github.com/Drifter-wu/PsiGWmodel}}.

The $\Psi_{\mathrm{FD}}$ model is derived from the Konoplya–Rezzolla–Zhidenko (KRZ) metric, which provides a model-independent framework for parameterizing generic black hole geometries using a finite number of tunable parameters. By appropriately adjusting these parameters, a variety of well-known black hole metrics, including the Kerr metric, can be exactly recovered over the entire spacetime. The metric takes the following form:
\begin{eqnarray}\label{metric}
d s^{2}&=&-\frac{N^{2}({r}, \theta)-W^{2}({r}, \theta) \sin ^{2} \theta}{K^{2}({r}, \theta)} d t^{2}\nonumber\\&&-2 W({r}, \theta) {r} \sin ^{2} \theta d t d \phi  \nonumber\\&&+K^{2}({r}, \theta) {r}^{2} \sin ^{2} \theta d \phi^{2}\nonumber\\&&+\Sigma({r}, \theta)\left(\frac{B^{2}({r}, \theta)}{N^{2}({r}, \theta)} d {r}^{2}+{r}^{2}d \theta^{2}\right), 
\end{eqnarray}

The other metric functions are defined as:

\begin{eqnarray}
\Sigma&=&1+a^{2} \cos ^{2} \theta/{r}^{2}\ ,\\
{N}^{2}&=&\left( 1-{{r}_{0}}/{r} \right)\nonumber\\&&\left[ 1-{{\epsilon }_{0}}{{r}_{0}}/{r}+\left( {{k}_{00}}-{{\epsilon }_{0}} \right)r_{0}^{2}/{{{{r}}}^{2}}+{{\delta }_{1}}r_{0}^{3}/{{{{r}}}^{3}} \right]\nonumber\\&&+[ {{a}_{20}}r_{0}^{3}/{{{{r}}}^{3}}+{{a}_{21}}r_{0}^{4}/{{{{r}}}^{4}}+{{k}_{21}}r_{0}^{3}/{{{{r}}}^{3}}L]{{\cos }^{2}}\theta\ ,\\
B&=&1+\delta_{4} r_{0}^{2} / {r}^{2}+\delta_{5} r_{0}^{2} \cos ^{2} \theta / {r}^{2}\ ,\\
W&=&\left[w_{00} r_{0}^{2} / {r}^{2}+\delta_{2} r_{0}^{3} / {r}^{3}+\delta_{3} r_{0}^{3} / {r}^{3} \cos ^{2} \theta\right] / \Sigma\ ,\\
K^{2}&=&1+a W / r\nonumber\\&&+\left\{k_{00} r_{0}^{2} / {r}^{2}+k_{21} r_{0}^{3} / {r}^{3}L \cos ^{2} \theta\right\} / \Sigma\ ,
\\ \label{L}
L&=&\left[1+\frac{k_{22}\left(1-r_{0} / {r}\right)}{1+k_{23}\left(1-r_{0} / {r}\right)}\right]^{-1}\ .
\end{eqnarray}

The parameters appearing in the above equations are defined as follows:
\begin{eqnarray}
a_{20}&=&2 \tilde{a}^{2} / r_{0}^{3}, a_{21} = -\tilde{a}^{4} / r_{0}^{4}+\delta_{6}, \\ \epsilon_{0}&=&\left(2-r_{0}\right) / r_{0}, k_{00} = \tilde{a}^{2} / r_{0}^{2} ,\\ 
k_{21}&=&\tilde{a}^{4} / r_{0}^{4}-2 \tilde{a}^{2} / r_{0}^{3}-\delta_{6}, \\ 
w_{00}&=& 2 \tilde{a} / r_{0}^{2}, k_{22} = -\tilde{a}^{2} / r_{0}^{2}+\delta_{7}, \\ 
k_{23}&=&\tilde{a}^{2} / r_{0}^{2}+\delta_{8},
\end{eqnarray}

where $r_0$ represents the equatorial radius of the event horizon, the dimensionless parameter $\delta_{i}$, where $i=1,2,3,4,5,6,7,8$, describes the deformation of various parameters in metric~(\ref{metric}). The detailed information of $\delta_{i}$ can be found in the KRZ paper.

The KRZ metric has the advantage of being relatively simple compared to other metrics, which motivates us to use it for constructing waveform templates. However, a limitation of the KRZ metric is that it cannot provide exact multipole moments unless it reduces to a specific metric. Fortunately, the bumpy black hole metric offers a more flexible multipolar structure that closely resembles that of standard black holes while allowing for controlled deviations. In the limit where the deviation parameters vanish, the bumpy black hole reduces to standard solutions such as the Schwarzschild or Kerr black hole. Therefore, we propose to use the bumpy black hole metric to obtain the exact multipole moments and then incorporate them into the KRZ framework, enabling the use of the $\Psi_{\mathrm{FD}}$ model to generate waveforms with additional multipolar. The bumpy Kerr black hole metric can be expressed in Boyer–Lindquist coordinates as follows:

\begin{equation}
\begin{aligned}
d s^2= & -e^{2 \psi_1}\left(1-\frac{2 M r}{\Sigma}\right) d t^2+\\
&e^{2 \psi_1-\gamma_1}\left(1-e^{\gamma_1}\right) \frac{4 a^2 M r \sin ^2 \theta}{\Delta \Sigma} d t d r \\
&-e^{2 \psi_1-\gamma_1} \frac{4 a M r \sin ^2 \theta}{\Sigma} d t d \phi +e^{2 \gamma_1-2 \psi_1}\left(1-\frac{2 M r}{\Sigma}\right)^{-1}\\
&[1+e^{-2 \gamma_1}\left(1-2 e^{\gamma_1}\right) \frac{a^2 \sin ^2 \theta}{\Delta}-\\
&e^{4 \psi_1-4 \gamma_1}\left(1-e^{\gamma_1}\right)^2 \frac{4 a^4 M^2 r^2 \sin ^4 \theta}{\Delta^2 \Sigma^2}] d r^2 \\
& -2\left(1-e^{\gamma_1}\right) a \sin ^2 \theta[e^{-2 \psi_1}\left(1-\frac{2 M r}{\Sigma}\right)^{-1}-\\
&e^{2 \psi_1-2 \gamma_1} \frac{4 a^2 M^2 r^2 \sin ^2 \theta}{\Delta \Sigma(\Sigma-2 M r)}] d r d \phi \\
& +e^{2 \gamma_1-2 \psi_1} \Sigma d \theta^2+\Delta [e^{-2 \psi_1}\left(1-\frac{2 M r}{\Sigma}\right)^{-1}-\\
&e^{2 \psi_1-2 \gamma_1} \frac{4 a^2 M^2 r^2 \sin ^2 \theta}{\Delta \Sigma(\Sigma-2 M r)}] \sin ^2 \theta d \phi^2
\end{aligned}
\end{equation}

The bumpy Kerr black hole metric can be expressed in the form $g_{\alpha\beta}=g^{\mathrm{Kerr}}_{\alpha\beta}+b_{\alpha\beta}$, where $g^{\mathrm{Kerr}}_{\alpha\beta}$ denotes the Kerr metric. In the above equation, $\Delta\equiv r^2-2Mr+a^2$, and $\gamma_1$ and $\psi_1$ denote the perturbation potentials arising from the mass moment and spin moment perturbations, respectively. The definitions of $\gamma_1$ and $\phi_1$ are detailed in \cite{Bumpy_BH_1, Bumpy_BH_2}. The bumpy Kerr black hole metric reduces to the Kerr black hole metric in the absence of perturbations, i.e., $\gamma_1=\phi_1=0$.

The perturbation of $\gamma_1$ and $\phi_1$ has an $l = 2$ spherical harmonic form in the Boyer-Lindquist coordinates\cite{Bumpy_BH_2}:

\begin{equation}
\begin{aligned}
\psi_1^{l=2}(r, \theta) & =\frac{B_2 M^3}{4} \sqrt{\frac{5}{\pi}} \frac{1}{d(r, \theta, a)^3}\left[\frac{3 L(r, \theta, a)^2 \cos ^2 \theta}{d(r, \theta, a)^2}-1\right], \\
\gamma_1^{l=2}(r, \theta) & =B_2 \sqrt{\frac{5}{\pi}}\\
[\frac{L(r, \theta, a)}{2} &\frac{\left[c_{20}(r, a)+c_{22}(r, a) \cos ^2 \theta+c_{24}(r, a) \cos ^4 \theta\right]}{d(r, \theta, a)^5}-1].\label{B2_first}
\end{aligned}
\end{equation}
where

\begin{equation}
\begin{aligned}
d(r, \theta, a)&=\sqrt{r^2-2 M r+\left(M^2+a^2\right) \cos ^2 \theta} \\
L(r, \theta, a)&=\sqrt{(r-M)^2+a^2 \cos ^2 \theta} \\
\end{aligned}
\end{equation}
and
\begin{equation}
\begin{aligned}
c_{20}(r, a)&=2(r-M)^4-5 M^2(r-M)^2+3 M^4, \\
c_{22}(r, a)&=5 M^2(r-M)^2-3 M^4+a^2\left[4(r-M)^2-5 M^2\right], \\
c_{24}(r, a)&=a^2\left(2 a^2+5 M^2\right) .
\end{aligned}
\end{equation}

By selecting the appropriate parameters, the KRZ metric can be reduced to the bumpy black hole metric. Since the KRZ metric does not provide an exact value for the black hole's quadrupole moment, we aim to utilize the quadrupole moment in the bumpy black hole metric to correspond to $\delta_i$. Upon performing these calculations, we found that choosing specific values for $\delta_i$ reduces the KRZ metric to the bumpy black hole metric:

\begin{equation}
\begin{aligned}
\delta_1 =& \left \{[(2\psi_1+1)(1-\frac{2M}{r})](1-\delta_6\frac{r_0^3}{r^3}\mathrm{cos^2\theta})-(1-\frac{r_0}{r} )  \right \}\\
&/[\frac{r_0^3}{r^3}(1-\frac{r_0}{r}  ) ],\\
\delta_6 = &\frac{2\psi_1r^5}{r_0^3}\mathrm{tan^2\theta},\\
\delta_2 =&\delta_3=\delta_4=\delta_5=0.
\end{aligned}
\end{equation}

The above equations give the quadrupole moment:
\begin{equation}
{Q}=-Ma^2-B_2M^3\sqrt{5/4\pi}={Q}_K+\Delta{Q} \label{DeltaQ_Eq}
\end{equation}
where $B_2$ is the parameter that appears in Eq.~(\ref{B2_first}). With Eq.~(\ref{DeltaQ_Eq}), we can construct the waveform template with the beyond-GR parameter $\Delta Q$. In the following context, we will briefly introduce the main differences of the $\Psi_{\mathrm{FD}}$ model.

\subsection{Inspiral}\label{Inspiral_sec}

During the inspiral phase, we construct the gravitational waveform within the effective one-body (EOB) framework. Imposing the normalization condition on the four-velocity, $u^{\mu}u_{\mu}=-1$, leads to:

\begin{equation}
V_{\mathrm{eff}}=g_{rr}\dot{r}^2=-1-g_{tt}\dot{t}^2-g_{\phi\phi}\dot{\phi}^2,
\end{equation}

To simplify the above equation, we can use the specific energy(the energy per unit mass) and specific angular momentum(the angular momentum per unit mass) of a particle:
\begin{eqnarray}
E=-(g_{tt}u^{t} +g_{t\phi }u^{\phi}),\label{energy}\\ 
L=g_{\phi t}u^{t} +g_{\phi\phi }u^{\phi}. \label{momentum}
\end{eqnarray}

Then, with the calculations, we can get the phase deformation for different deformation parameters (i.e., $\delta_i$). For example, the phase deformation due to $\delta_1$ is:
\begin{equation}
\phi^{\delta_1}_{\mathrm{KRZ}} = -\frac{75}{8}u^{-1/3}\eta^{-4/5}\delta_1
\end{equation}
where $u=\eta \pi mf$ and $\eta=\frac{m_1m_2}{m_1+m_2}$ is symmetric mass ratio. The other additional phase could be found in \cite{Li_23}.

Then we can get the phase ansatz in the inspiral stage: 

\begin{equation}
\begin{aligned}
\phi_{\mathrm{Ins}}= & \phi_{\mathrm{TF} 2}(M f ; \Xi) \\
& +\frac{1}{\eta}\left(\sigma_{0}+\sigma_{1} f+\frac{3}{4} \sigma_{2} f^{4 / 3}+\frac{3}{5} \sigma_{3} f^{5 / 3}+\frac{1}{2} \sigma_{4} f^{2}\right)\\
&+\phi_{\mathrm{KRZ}}
\end{aligned}
\end{equation}

where $\eta=m_1 m_2/M^2$, $M = m_1+m_2$, the $\phi_{\mathrm{TF} 2}$ is the full TaylorF2 phase:
\begin{equation}\label{Ins_equ_D}
\begin{aligned}
\phi_{\mathrm{TF} 2}= & 2 \pi f t_{c}-\varphi_{c}-\pi / 4 \\
& +\frac{3}{128 \eta}(\pi f M)^{-5 / 3} \sum_{i=0}^{7} \varphi_{i}(\Xi)(\pi f M)^{i / 3}
\end{aligned}
\end{equation}
The constants $\sigma_{i}$ (where $i = 0, 1, 2, 3, 4$) represent the correlation between the mass and spin of the system. Meanwhile, the phase deformation arising from the general parameterized black hole is denoted by $\phi_{\mathrm{KRZ}}$. Varying the values of $\delta_1$, $\delta_2$, $\delta_4$, and $\delta_6$ will result in different phases. $\varphi_{i}(\Xi)$ are the PN expansion coefficients that are related to the intrinsic binary parameters. The detailed information of $\sigma_{i}$ and $\varphi_{i}(\Xi)$ can be found in Appendix B of the article\cite{PhenomD}. Because the duration of the intermediate is indeed short, we adopt the same prescription for the intermediate-phase evolution as that used in PhenomD.

\subsection{Ringdown}\label{Ringdown_sec}
The ringdown waveform is modeled analytically based on photon motion near the black hole photon sphere. Within this framework, the quasinormal modes of the remnant black hole are determined by the properties of unstable circular null geodesics. Specifically, the real part of the QNM frequency is associated with the fundamental frequencies of the photon orbit, while the imaginary part is determined by the Lyapunov exponent that characterizes the orbital instability. In Ref.~\cite{McWilliams_19}, the authors proposed the Backwards One-Body (BOB) method, which contains no phenomenological free parameters yet achieves an accuracy comparable to that of the most precise existing models. The ringdown waveform can then be written as

\begin{equation}\label{RD_equ_D}
{h}_{22}=X \operatorname{sech}\left[\gamma\left(t-t_{p}\right)\right] e^{-i \tilde{\Phi}_{22}(t)}\,,
\end{equation}

where $X$ is a constant related to the amplitude of the waveform, $\gamma$ is the Lyapunov exponent characterizing the rate of divergence of nearby null geodesics, $t_p$ is the time at maximum amplitude of the waveform, and $\tilde{\Phi}_{22}(t)$ is the phase. Both $\gamma$ and $\tilde{\Phi}_{22}(t)$ are related to the QNMs. Therefore, the parametrized deviations $\delta_i$ from the KRZ metric modify the photon sphere properties and consequently shift both the oscillation frequencies and damping rates of the QNMs, thereby affecting the ringdown waveform. As a result, the deformation parameters $\delta_i$ in the KRZ metric lead to a ringdown waveform that differs from that of the Kerr spacetime.

\begin{figure*}
\centering
\includegraphics[width=1.00 \textwidth]{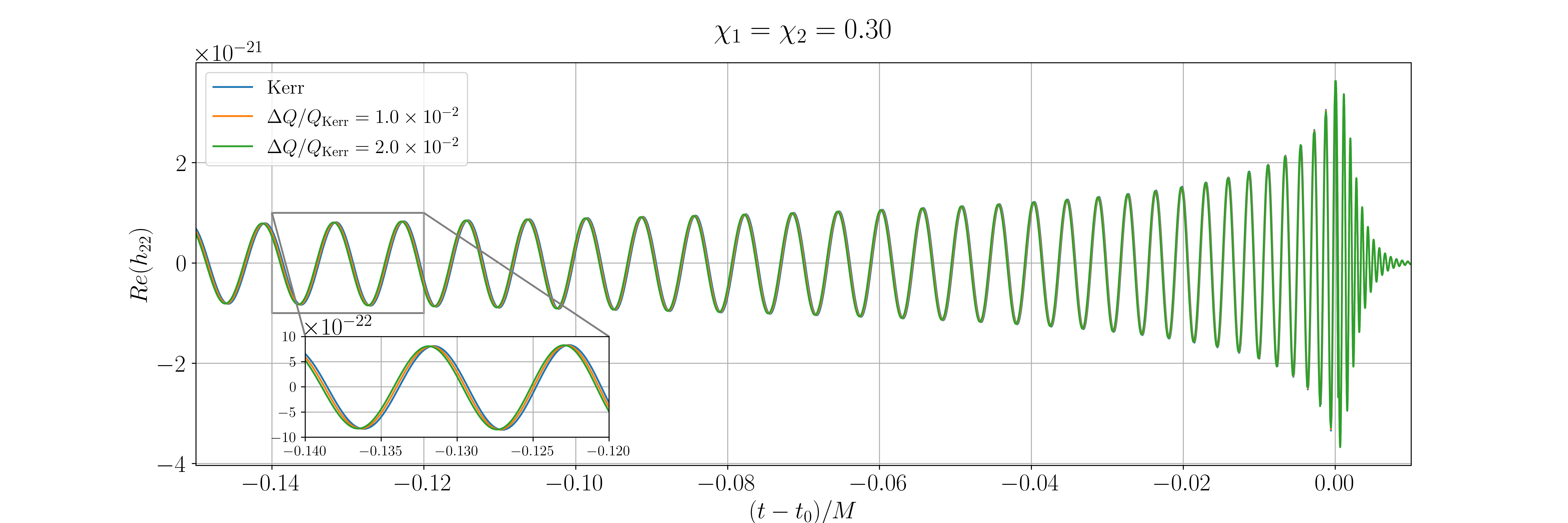}
\includegraphics[width=1.00 \textwidth]{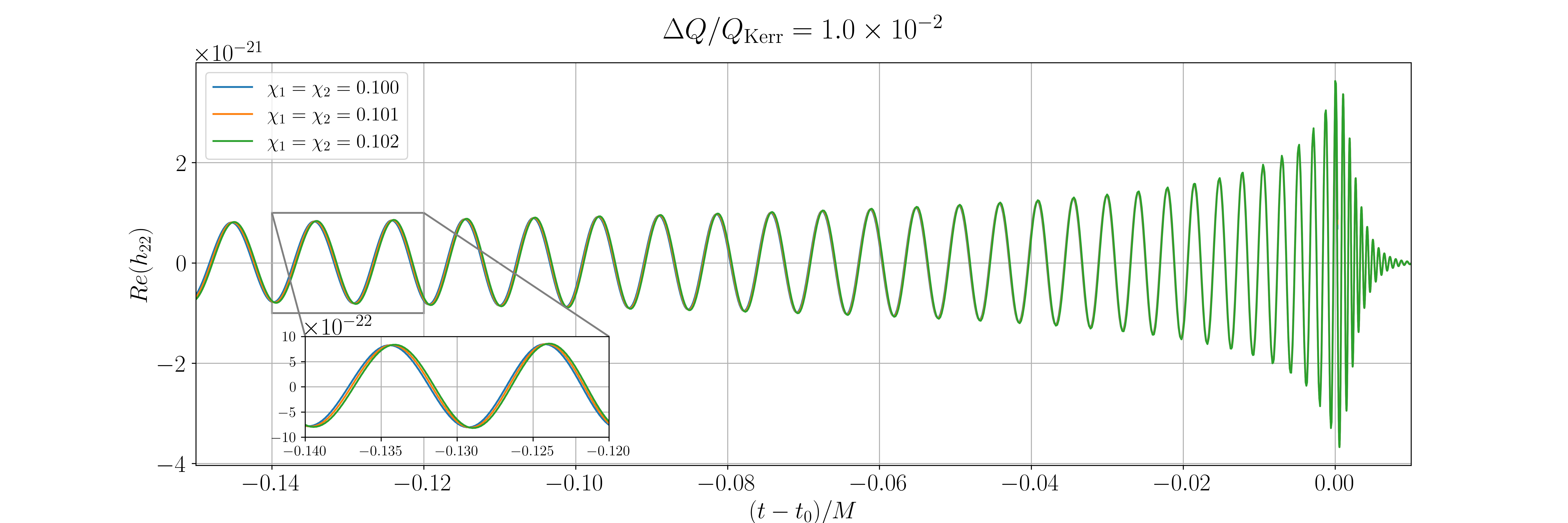}
\caption{The waveform of the $\Psi_{\mathrm{FD}}$ model. The top panel shows the different $\Delta Q/Q_{\mathrm{Kerr}}$ at the same spin $\chi_1=\chi_2=0.30$, the bottom panel shows the different spins at the same $\Delta Q/Q_{\mathrm{Kerr}}$. All the cases consider the same mass ratio 1:1. }\label{Psi_figure}
\end{figure*}

With the above context, we can get the full waveform $\Psi_{\mathrm{FD}}$. The more detailed study of the $\Psi_{\mathrm{FD}}$ waveform model can be found in \cite{Li_23}. In Fig.~\ref{Psi_figure}, we show the $\Psi_{\mathrm{FD}}$ waveform model for different values of $\Delta Q/Q_{\mathrm{Kerr}}$ at fixed spin in the top panel, and for different spins at fixed$\Delta Q/Q_{\mathrm{Kerr}}$ in the bottom panel. From the top panel, we see that as $\Delta Q/Q_{\mathrm{Kerr}}$ increases, the waveform departs progressively further from the Kerr waveform. In the bottom panel, we find that even with $\Delta Q/Q_{\mathrm{Kerr}}$ fixed, small variations in the spin still lead to noticeable differences in the waveform.

\section{Parameter estimation}\label{PE_sec}
To extract the physical properties of compact binary systems from gravitational-wave observations, we perform Bayesian parameter estimation using the Python-based software package \texttt{Bilby} which provides a flexible framework for performing inference with a variety of samplers and waveform templates, allowing robust estimation of source parameters from observed strain data \cite{bilby_paper}.

\begin{table}[]
\setlength{\tabcolsep}{5mm}{
\begin{tabular}{llll}
\hline\hline
Variable                 & Unit        & Prior   & Range             \\ \hline
$m_{1, 2}$         & $M_{\odot}$ & Uniform & (20, 55)          \\
$D_L$                    & Mpc         & Uniform & (200, 1250)       \\
$\Delta Q/Q$             & ...         & Uniform & (-0.10, 0.10)      \\ \hline
$a_{1, 2}$               & ...         & Uniform & (0, 0.99)         \\
$\theta_{1, 2}$          & rad         & sin     & (0, $\pi$)        \\
$\delta \phi, \phi_{\mathrm{JL}}$ & rad         & Uniform & (0, 2$\pi$)       \\
R.A.                     & rad         & Uniform & (0, 2$\pi$)       \\
Decl.                     & rad         & cos     & ($-\pi/2, \pi/2$) \\
$\iota$                  & rad         & sin     & (0, $\pi$)        \\
$\psi$                   & rad         & Uniform & (0, $\pi$)        \\
$\phi_c$                 & rad         & Uniform & (0, 2$\pi$)       \\ \hline
\end{tabular}
}
\caption{\textbf{Prior Setting.} The prior ranges of all the parameters for the gravitational wave events. Except mass $m_{1, 2}$, luminosity distance $D_L$, and relative deviation quadrupole moment $\Delta Q/Q$, other parameters are consistent with Table 1 in \cite{bilby_paper}.}
\label{Priors}
\end{table}

In the Bayesian framework, the posterior distribution of the parameters $\boldsymbol{\theta}$ given the observed data $d$ and a waveform template $\mathcal{H}$ is expressed as:

\begin{equation}
p(\boldsymbol{\theta} \mid d, \mathcal{H}) \propto \mathcal{L}(d \mid \boldsymbol{\theta}, \mathcal{H}) \, \pi(\boldsymbol{\theta}),
\end{equation}

where $\mathcal{L}$ is the likelihood function and $\pi(\boldsymbol{\theta})$ represents the prior distribution of the parameters. \texttt{Bilby} provides a flexible framework for defining custom likelihoods and priors, and for performing posterior sampling with different samplers. The parameter vector in this work is:

\begin{equation}
\begin{aligned}
\boldsymbol{\theta} = \bigl(&
m_1,m_2,D_L,\Delta Q/{Q},a_1,a_2,\theta_1,\theta_2,\delta\phi, \\
& \phi_{\mathrm{JL}},\mathrm{R.A.},\mathrm{Decl.},\iota,\psi,\phi_c
\bigr)
\end{aligned}
\end{equation}

Here, $\boldsymbol{\theta}$ denotes the set of source parameters. Specifically, $m_1$ and $m_2$ are the component masses of the binary, $D_L$ is the luminosity distance, and $\Delta Q / Q$ characterizes the fractional deviation of the quadrupole moment from the Kerr value. The quantities $a_1$ and $a_2$ denote the dimensionless spin magnitudes, while $\theta_1$ and $\theta_2$ are the tilt angles between the individual spin vectors and the orbital angular momentum. The parameter $\delta\phi$ is the relative azimuthal angle between the two spin vectors, and $\phi_{\mathrm{JL}}$ is the azimuthal angle describing the orientation of the total angular momentum. Furthermore, $\mathrm{R.A.}$ and $\mathrm{Decl.}$ represent the right ascension and declination of the source, respectively, $\iota$ is the inclination angle between the line of sight and the orbital angular momentum, $\psi$ is the polarization angle, and $\phi_c$ is the coalescence phase.

Table.~\ref{Priors} shows the prior setting for the analysis in this work. If we want to compare which gravitational waveform templates are better at describing events, we should use the Bayes factor:
\begin{equation}
\mathrm{BF}_{B}^{A}=\frac{\mathcal{Z}_{A}}{\mathcal{Z}_{B}}
\end{equation}

where $\mathcal{Z}_{A}$, $\mathcal{Z}_{B}$ are the Bayesian evidences of models A and B.

The definition of the Bayesian evidence is:

\begin{equation}
\mathcal{Z}=\mathrm{p}(d \mid \mathcal{H})=\int \mathrm{p}(\boldsymbol{\theta} \mid \mathcal{H}) \mathrm{p}(d \mid \boldsymbol{\theta}, \mathcal{H}) \mathrm{d} \boldsymbol{\theta}
\end{equation}

In this work, we compare the $\Psi_{\mathrm{FD}}$ waveform template with the \texttt{IMRPhenomXPHM} waveform template and then calculate the Bayes factor $\mathrm{BF}$ to see which waveform template is better to describe the gravitational wave events. To enhance the data’s intuitiveness, we use the common logarithm of the Bayes factor $\left |\mathrm{log} \, \mathrm{BF}  \right |$. As discussed in Refs.~\cite{Thrane_2019, Robert_95}, a larger value of $\log \mathrm{BF}$ indicates stronger evidence in favor of one model relative to another. Generally, a threshold of $\left |\mathrm{log} \, \mathrm{BF}  \right |  =8$ is often used as the level of ``very strong evidence'' in favor of one model over another, and the value in $\left [6, 8\right )$, $\left [2, 6\right )$ and $\left [0, 2\right )$, provide ``strong evidence'', ``positive'', and barely worth mentioning. In previous work, we found that only events with high SNRs yield large values of $\left |\mathrm{log} \, \mathrm{BF}  \right |$. Accordingly, in this work, we mainly focus on high-SNR events from the O4 observing run.

\subsection{Inspiral and post-inspiral decomposition}
\label{InspPostInsp_sec}
Motivated by the inspiral and post-inspiral tests developed in Refs.~\cite{Ghosh2016, Ghosh2018, LIGO2016TestsGR, LIGO2021TestsGR, LIGO2026TestsGR}, we further investigate the relative quadrupole moment deviation $\Delta Q/Q$ by analyzing the inspiral and post-inspiral portions of the gravitational-wave signal separately. In the conventional consistency test, the low- and high-frequency portions of a binary black hole signal are analyzed independently, and the corresponding estimates of the remnant mass and spin are compared to test the internal consistency of general relativity. Here, we adopt the same decomposition and apply it directly to the quadrupole moment deviation parameter $\Delta Q/Q$ within the $\Psi_{\mathrm{FD}}$ framework.

The inspiral and post-inspiral portions of a binary black hole signal probe different dynamical regimes of the coalescence and may therefore contain complementary information about deviations from the Kerr quadrupole moment. During the inspiral, the quadrupole moment affects the orbital dynamics. Therefore, the signal can contain many inspiral cycles, and even a relatively small modification may produce a significant phase correction. By contrast, the post-inspiral portion probes the highly nonlinear merger and the subsequent relaxation of the remnant, where the multipolar structure can leave distinct imprints on the waveform.

We therefore perform independent parameter-estimation analyses using the inspiral and post-inspiral frequency ranges and compare the resulting posterior distributions of $\Delta Q/Q$. This decomposition allows us to determine which portion of the signal provides the dominant constraint on the quadrupole moment deviation and to assess the robustness of the corresponding full-waveform result.

\begin{figure*}
\centering
\includegraphics[width=1.00 \textwidth]{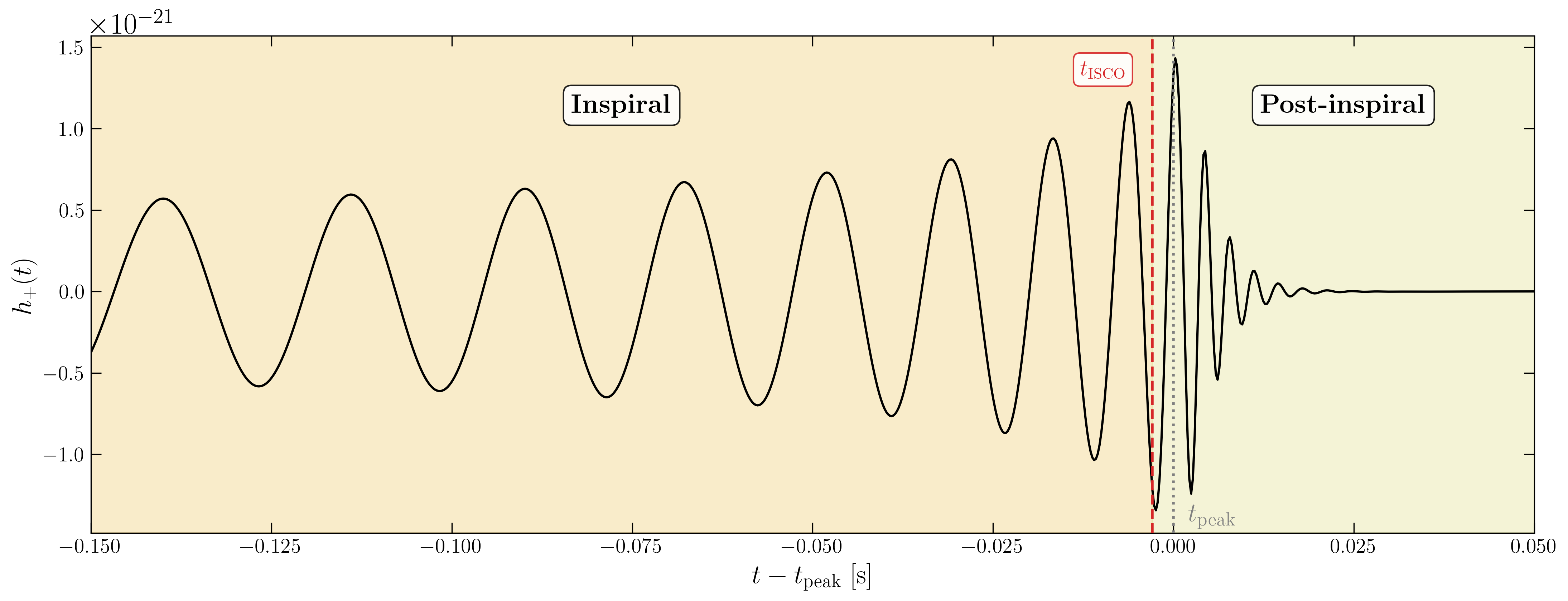}
\caption{Schematic time-domain gravitational waveform illustrating the separation between the inspiral and post-inspiral regimes. The red dashed vertical line marks the time $t_{\rm ISCO}$ associated with the cutoff frequency $f_{\rm cut}$, defined from the innermost stable circular orbit of the remnant Kerr black hole. The gray dotted vertical line denotes the peak of the waveform amplitude. The inspiral regime corresponds to $t<t_{\rm ISCO}$, whereas the post-inspiral regime begins at $t_{\rm ISCO}$ and includes the late inspiral, merger, and ringdown. This time-domain representation is schematic, since the parameter-estimation analyses are separated using the corresponding cutoff frequency in the frequency domain.}
\label{insp_postinsp_fig}
\end{figure*}

We separate the two regimes by introducing an event-dependent cutoff frequency $f_{\mathrm{cut}}$,
\begin{equation}
    f < f_{\mathrm{cut}} \quad \mathrm{(inspiral)}, \qquad
    f \geq f_{\mathrm{cut}} \quad \mathrm{(post\mbox{-}inspiral)} .
\end{equation}
We define the cutoff frequency as the dominant-mode gravitational-wave frequency associated with the innermost stable circular orbit (ISCO) of the remnant Kerr black hole\cite{Ghosh2018}. Fig.~\ref{insp_postinsp_fig} provides a schematic time-domain representation of the inspiral--post-inspiral decomposition.



\section{Results}\label{Results_sec}
\begin{figure*}
\centering
\includegraphics[width=1.00 \textwidth]{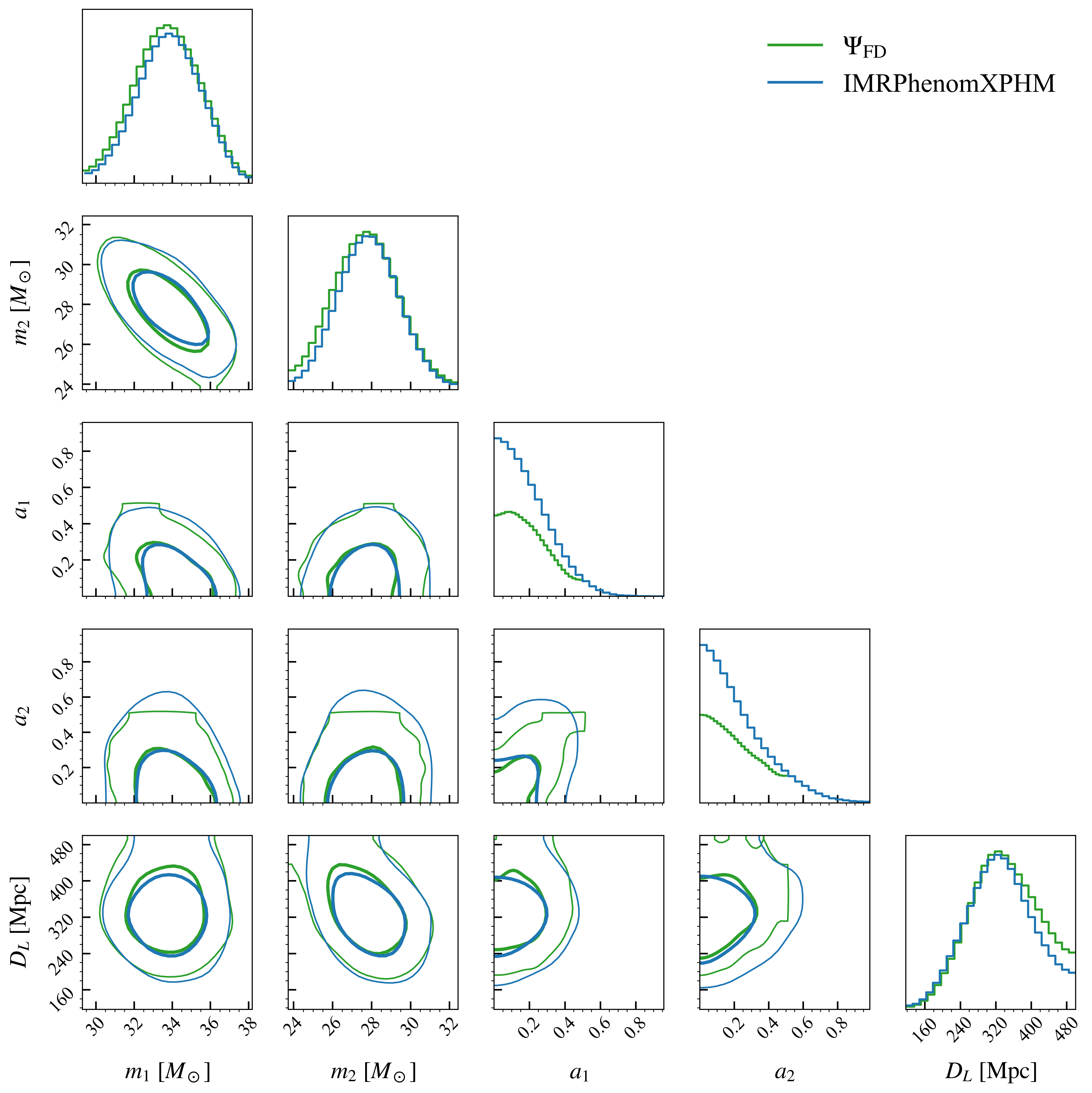}
\caption{Posterior distributions on selected properties of GW230814 for $\Psi_{\mathrm{FD}}$ (green) and \texttt{IMRPhenomXPHM} (blue): their masses $m_1$ and $m_2$, spins $a_1$ and $a_2$, distance $D_L$. Panels along the diagonal show marginalized posteriors on each parameter. Off-diagonal panels illustrate joint two-dimensional posteriors on each pair of parameters; thick and thin contours denote central 50\% and 90\% credible bounds.}\label{PE_fig1}
\end{figure*}

\begin{figure*}
\centering
\includegraphics[width=1.00 \textwidth]{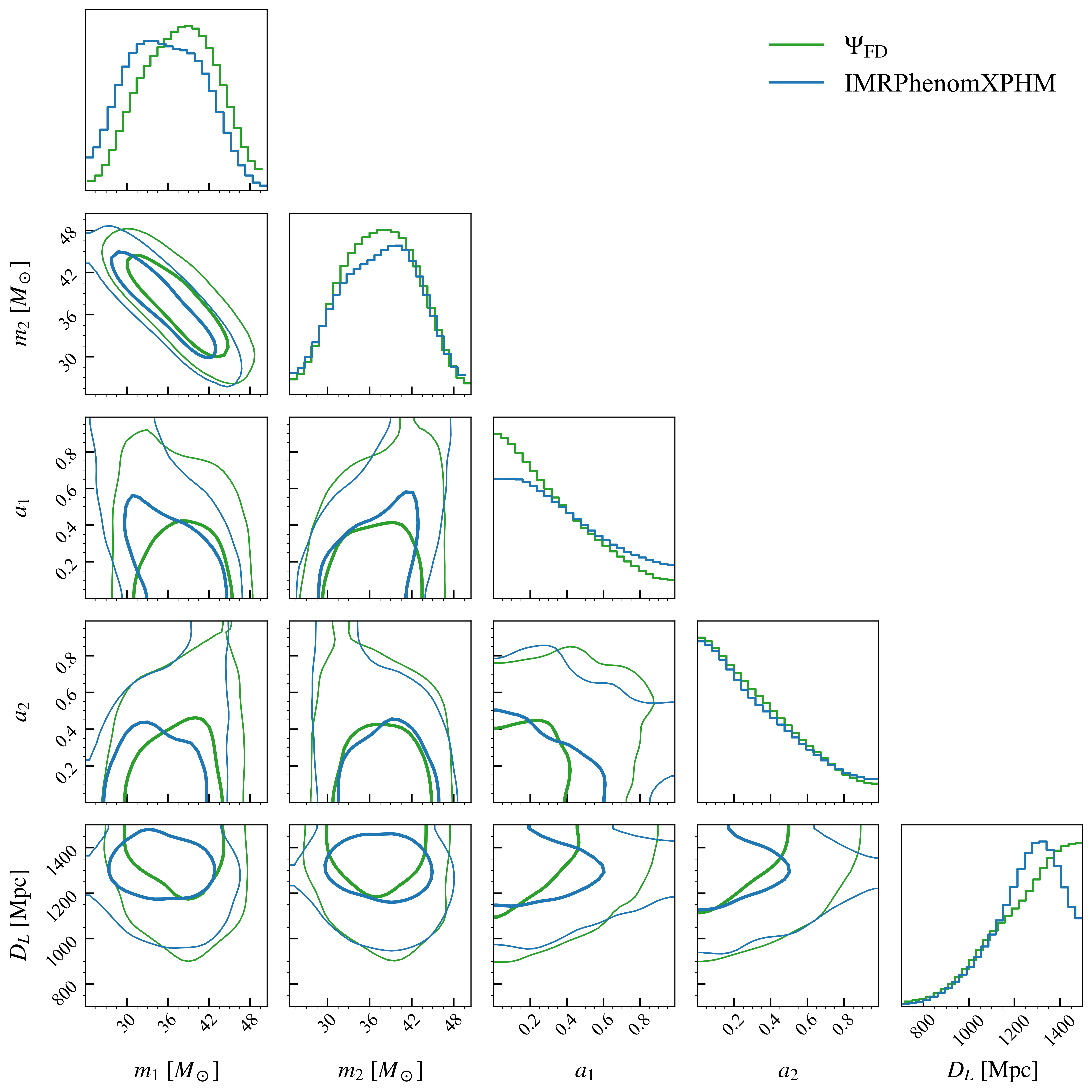}
\caption{Posterior distributions on selected properties of GW231226 for $\Psi_{\mathrm{FD}}$ (green) and \texttt{IMRPhenomXPHM} (blue): their masses $m_1$ and $m_2$, spins $a_1$ and $a_2$, distance $D_L$. Panels along the diagonal show marginalized posteriors on each parameter. Off-diagonal panels illustrate joint two-dimensional posteriors on each pair of parameters; thick and thin contours denote central 50\% and 90\% credible bounds.}\label{PE_fig2}
\end{figure*}

\begin{figure*}
\centering
\includegraphics[width=1.00 \textwidth]{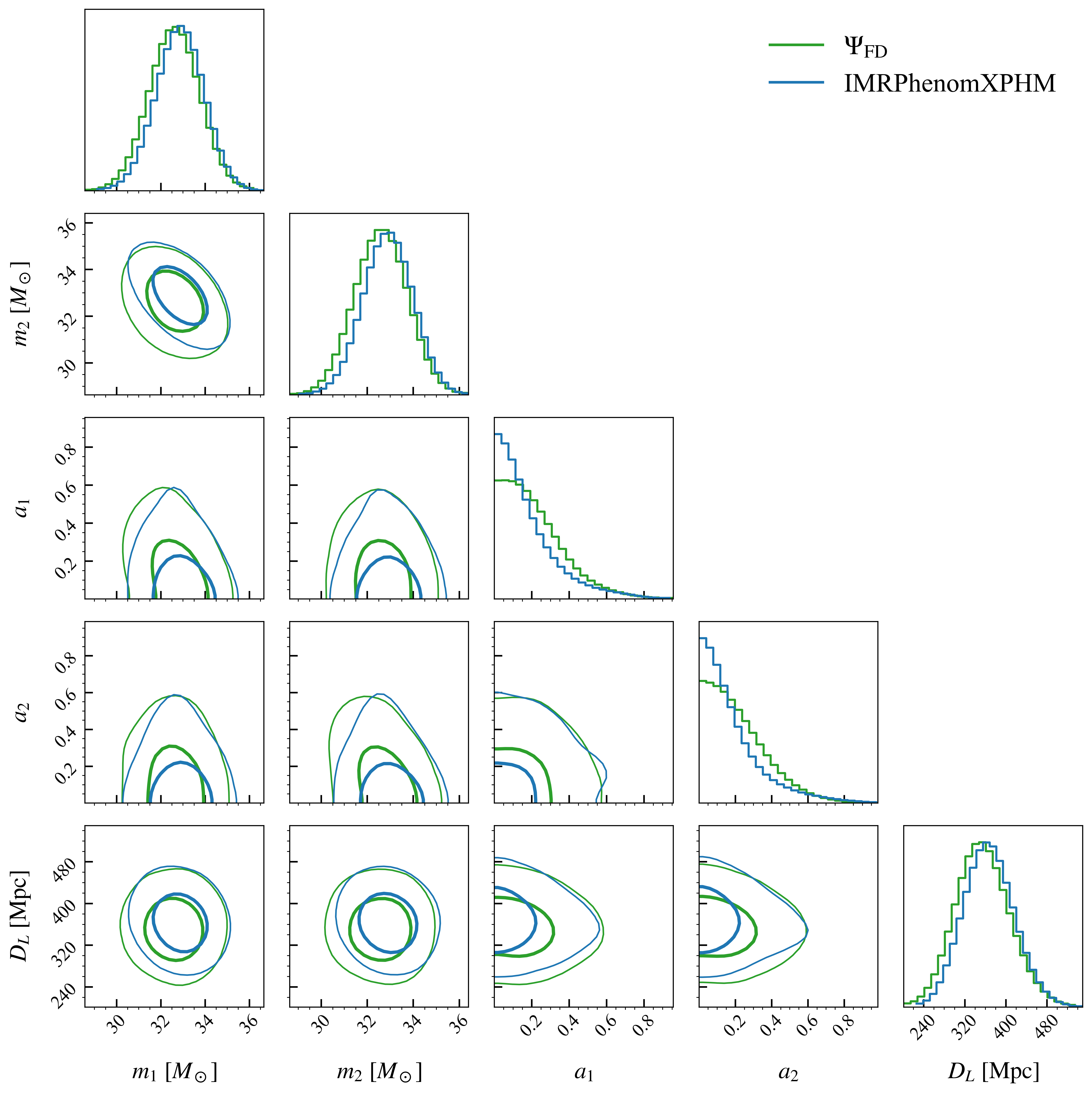}
\caption{Posterior distributions on selected properties of GW250114 for $\Psi_{\mathrm{FD}}$ (green) and \texttt{IMRPhenomXPHM} (blue): their masses $m_1$ and $m_2$, spins $a_1$ and $a_2$, distance $D_L$. Panels along the diagonal show marginalized posteriors on each parameter. Off-diagonal panels illustrate joint two-dimensional posteriors on each pair of parameters; thick and thin contours denote central 50\% and 90\% credible bounds.}\label{PE_fig3}
\end{figure*}

Figs.~\ref{PE_fig1}-\ref{PE_fig3} compare the posterior distributions obtained from full signal using $\Psi_{\mathrm{FD}}$ and $\texttt{IMRPhenomXPHM}$ for GW230814, GW231226, and GW250114, respectively. The corresponding comparisons for GW150914 and GW200129 were presented in Ref.~\cite{Li_24}. In each figure, we show the posterior distributions for a representative set of intrinsic and extrinsic source parameters, namely the component masses \((m_1, m_2)\), the dimensionless spin magnitudes \((a_1, a_2)\), and the luminosity distance \(D_L\). The diagonal panels display the one-dimensional marginalized posteriors, while the off-diagonal panels present the corresponding two-dimensional joint posteriors, with the inner and outer contours indicating the \(50\%\) and \(90\%\) credible regions. Overall, the posteriors inferred with \(\Psi_{\mathrm{FD}}\) are in good agreement with those obtained using \(\texttt{IMRPhenomXPHM}\), indicating that the two waveform models yield broadly consistent constraints on the source parameters for both events. The small differences between the two sets of posteriors may be attributed to the inclusion of the additional parameter \(\Delta Q/Q\) in \(\Psi_{\mathrm{FD}}\), which can induce slight shifts in the posterior contour shapes. 

\begin{figure*}
\centering
\includegraphics[width=0.80 \textwidth]{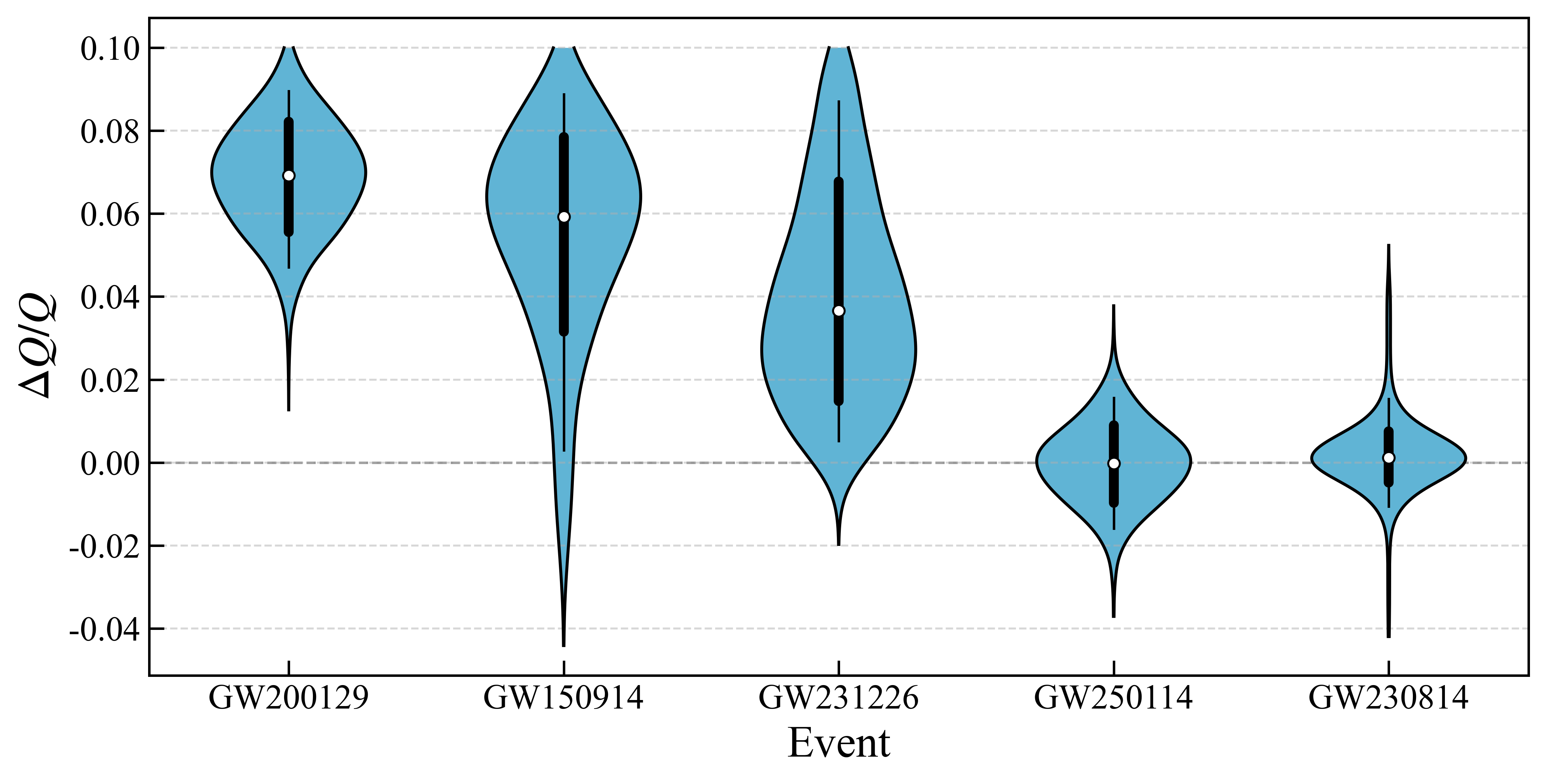}
\caption{Posterior distributions of $\Delta Q/Q$ for the different events inferred with the $\Psi_{\mathrm{FD}}$ model. The violin shapes show the kernel density estimates of the posterior samples of $\Delta Q/Q$ for each event. The white dot denotes the median, while the thick and thin vertical lines mark the 50\% and 90\% credible intervals, respectively. The dashed line at $\Delta Q/Q=0$ denotes the GR prediction.}\label{DeltaQ_Q_fig}
\end{figure*}

Fig.~\ref{DeltaQ_Q_fig} shows the posterior distributions of $\Delta Q/Q$ for the five different events from the full signal. To facilitate comparison, we arrange these events according to the inferred relative quadrupole moment deviation $\Delta Q/Q$, rather than in chronological order. The results indicate that GW200129 and GW150914 favor a nonzero relative quadrupole moment deviation $\Delta Q/Q$. The corresponding SNRs are 28.4 and 25.1, respectively. GW150914 exhibits a relatively weak constraint on $\Delta Q/Q$, possibly because it has the lowest SNR among the five events. By contrast, the posterior distributions for GW250114 and GW230814 are both peaked near $\Delta Q/Q=0$. We then investigated whether the posterior distributions of $\Delta Q/Q$ for GW250114 and GW230814 could be due to the $\Psi_{\mathrm{FD}}$ waveform model being less favored for these events than $\texttt{IMRPhenomXPHM}$. To this end, we computed the Bayes factors between the two waveform models and obtained $\mathrm{logBF} =-1.21$ and $0.04$ for GW250114 and GW230814, respectively. 
Since the absolute values of the logarithmic Bayes factors for both events lie in the interval $\left [0, 2\right )$, the model comparison provides only evidence that is barely worth mentioning according to the criterion introduced above. Thus, the data do not show a statistically significant preference for either waveform model. We therefore infer that, for GW250114 and GW230814, the relative quadrupole moment deviation $\Delta Q/Q$ is likely close to zero.

Table.~\ref{Events_diff_Table} summarizes the parameter-estimation results for the source-frame component masses $m_1$ and $m_2$, the luminosity distance $D_L$, and the relative quadrupole moment deviation $\Delta Q/Q$. The SNR denotes the signal-to-noise ratio, while the final column reports the common logarithm of the Bayes factor comparing the $\Psi_{\mathrm{FD}}$ and \texttt{IMRPhenomXPHM} waveform models. For the high-SNR events considered here, the inferred values of $\Delta Q/Q$ remain close to zero. 

\begin{table*}
    \renewcommand{\arraystretch}{1.5}
    \setlength{\tabcolsep}{12pt}
    \centering
    \begin{tabular}{|c|c|c|c|c|c|c|} \hline Events & $m_1$ & $m_2$ & $D_L$ & SNR & $\Delta Q/Q$ & $\log \mathrm{BF}$ \\ \hline GW200129 & $32.1^{+3.0}_{-4.2}$ & $26.5^{+4.4}_{-2.4}$ & $988^{+80}_{-91}$ & 28.4 & $0.07^{+0.02}_{-0.02}$ & 5.10 \\ \hline GW150914 & $34.8^{+4.9}_{-4.3}$ & $33.0^{+3.7}_{-3.7}$ & $447^{+99}_{-119}$ & 25.1 & $0.06^{+0.03}_{-0.05}$ & 1.59 \\ \hline GW231226 & $37.9^{+7.7}_{-8.2}$ & $37.5^{+7.8}_{-7.9}$ & $1309^{+171}_{-338}$ & 34.9 & $0.04^{+0.02}_{-0.03}$ & 0.62 \\ \hline GW250114 & $32.6^{+1.6}_{-1.6}$ & $32.6^{+1.6}_{-1.6}$ & $353^{+84}_{-70}$ & 80.0 & $0.00^{+0.01}_{-0.01}$ & -1.21 \\ \hline GW230814 & $33.7^{+2.6}_{-2.7}$ & $27.7^{+2.6}_{-2.5}$ & $337^{+132}_{-113}$ & 40.3 & $0.00^{+0.01}_{-0.01}$ & 0.04 \\ \hline \end{tabular}
    \caption{The table lists the median values and symmetric 90\% credible intervals for selected source parameters from our parameter-estimation analysis. Here $m_1$ and $m_2$ are the source-frame component masses in units of $M_\odot$, $D_L$ denotes the luminosity distance in Mpc, SNR is the signal-to-noise ratio, $\Delta Q/Q$ is the relative quadrupole moment deviation, and $\log \mathrm{BF}$ is the common logarithm of the Bayes factor comparing the $\Psi_{\mathrm{FD}}$ model with the Kerr model. Detailed posterior distributions for GW150914 and GW200129 were presented in our previous work~\cite{Li_24}.}
    \label{Events_diff_Table}
\end{table*}

\begin{figure}
\centering
\includegraphics[width=\columnwidth]{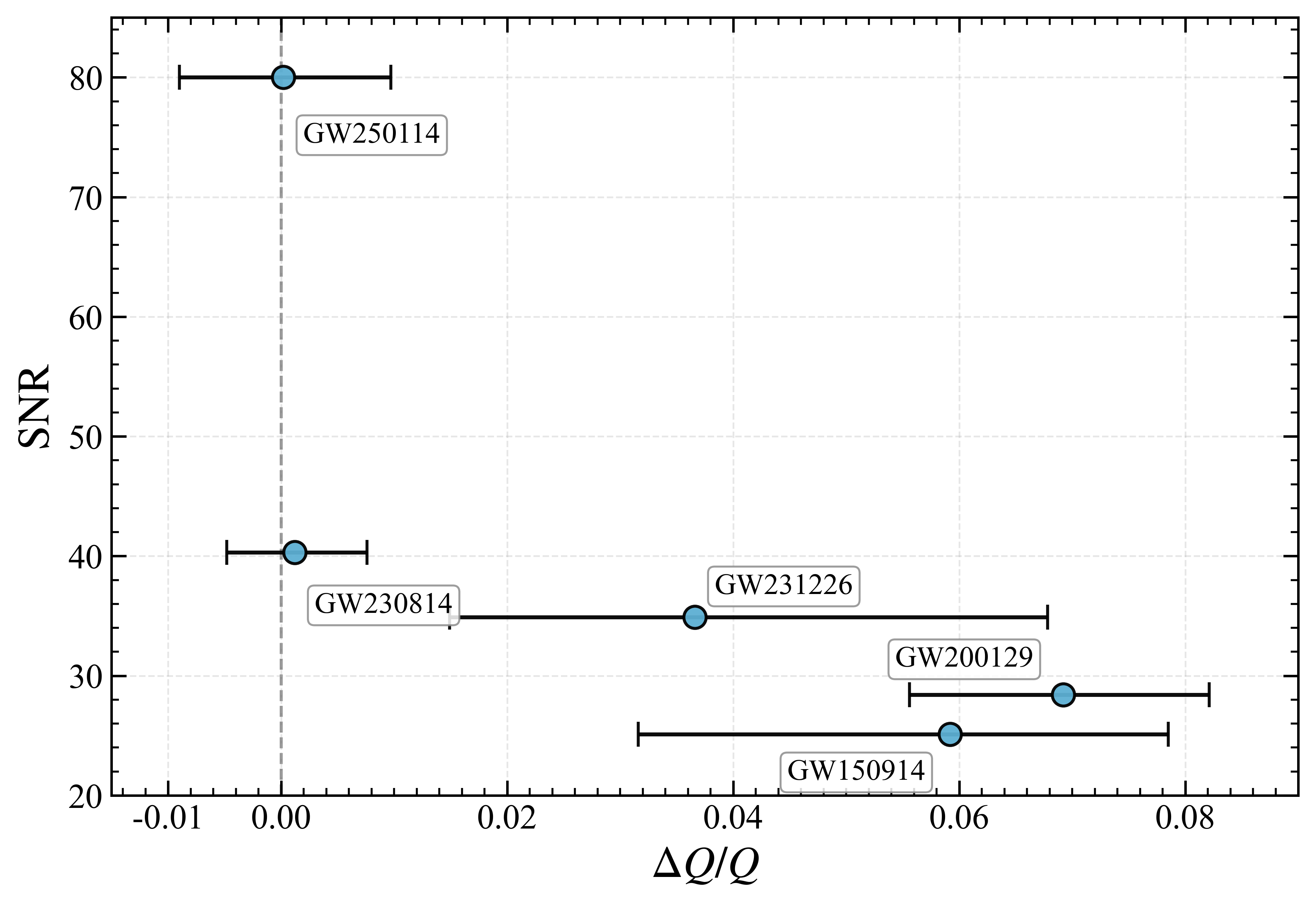}
\caption{Comparison of the inferred relative quadrupole moment deviations, $\Delta Q/Q$, and the signal-to-noise ratios (SNRs) for different gravitational-wave events. The markers indicate the posterior medians of $\Delta Q/Q$, and the horizontal error bars represent the corresponding $1\sigma$ credible intervals. The vertical dashed line at $\Delta Q/Q=0$ denotes the GR prediction.}\label{SNR_DeltaQ}
\end{figure}

We further compare the relative quadrupole moment deviations, $\Delta Q/Q$, with the signal-to-noise ratios of the analyzed events in Fig.~\ref{SNR_DeltaQ}. For the events considered here, those with higher SNRs tend to yield values of $\Delta Q/Q$ closer to zero. By contrast, GW150914 and GW200129, which were analyzed in our previous work~\cite{Li_24}, exhibit comparatively larger nonzero deviations. This behavior may be partly related to their lower SNRs, which lead to weaker constraints on $\Delta Q/Q$.

An interesting feature of Fig.~\ref{SNR_DeltaQ} is the difference between GW230814 and GW231226, despite their relatively similar SNRs of $40.3$ and $34.9$, respectively. The relative quadrupole moment deviations for the two events show different behaviors, with GW230814 yielding a value consistent with zero, while GW231226 favors a nonzero deviation. This difference may be associated with the smaller effective spin of GW230814. In low-spin systems, the spin-induced quadrupolar imprint on the waveform is expected to be weaker, which may reduce the sensitivity to quadrupole moment deviations. 

We then perform separate parameter-estimation analyses using the inspiral and post-inspiral portions of each signal. The cutoff frequencies adopted for the different events are listed in Tab.~\ref{fcut_table}. Because the results for GW150914 and GW200129 show qualitative similarities to those for GW231226, particularly in their nonzero relative quadrupole moment deviations and moderate SNRs, we show GW231226 as a representative example and do not display the corresponding analyses for GW150914 and GW200129.

\begin{table}
\centering
{
\setlength{\tabcolsep}{8pt}
\begin{tabular*}{\columnwidth}{
    @{\extracolsep{\fill}}lcccc
}
\hline\hline
Event 
& $f_{\mathrm{cut}}$ (Hz)
& $\rho_{\mathrm{Full}}$
& $\rho_{\mathrm{I}}$
& $\rho_{\mathrm{PI}}$ \\ 
\hline
GW231226 & 102 & 35 & 25.9 & 20.5 \\
GW230814 & 144 & 40 & 33.5 & 22.5 \\
GW250114 & 135 & 75 & 57.0 & 35.0 \\
\hline
\end{tabular*}
}
\caption{
$f_{\mathrm{cut}}$ represents cutoff frequencies between the inspiral and post-inspiral regimes. $\rho_{\mathrm{Full}}$, $\rho_{\mathrm{I}}$, and $\rho_{\mathrm{PI}}$  are the SNR in the full signal, the inspiral part, and the post-inspiral part, respectively.}
\label{fcut_table}
\end{table}

\begin{figure*}
\centering
\includegraphics[width=0.80 \textwidth]{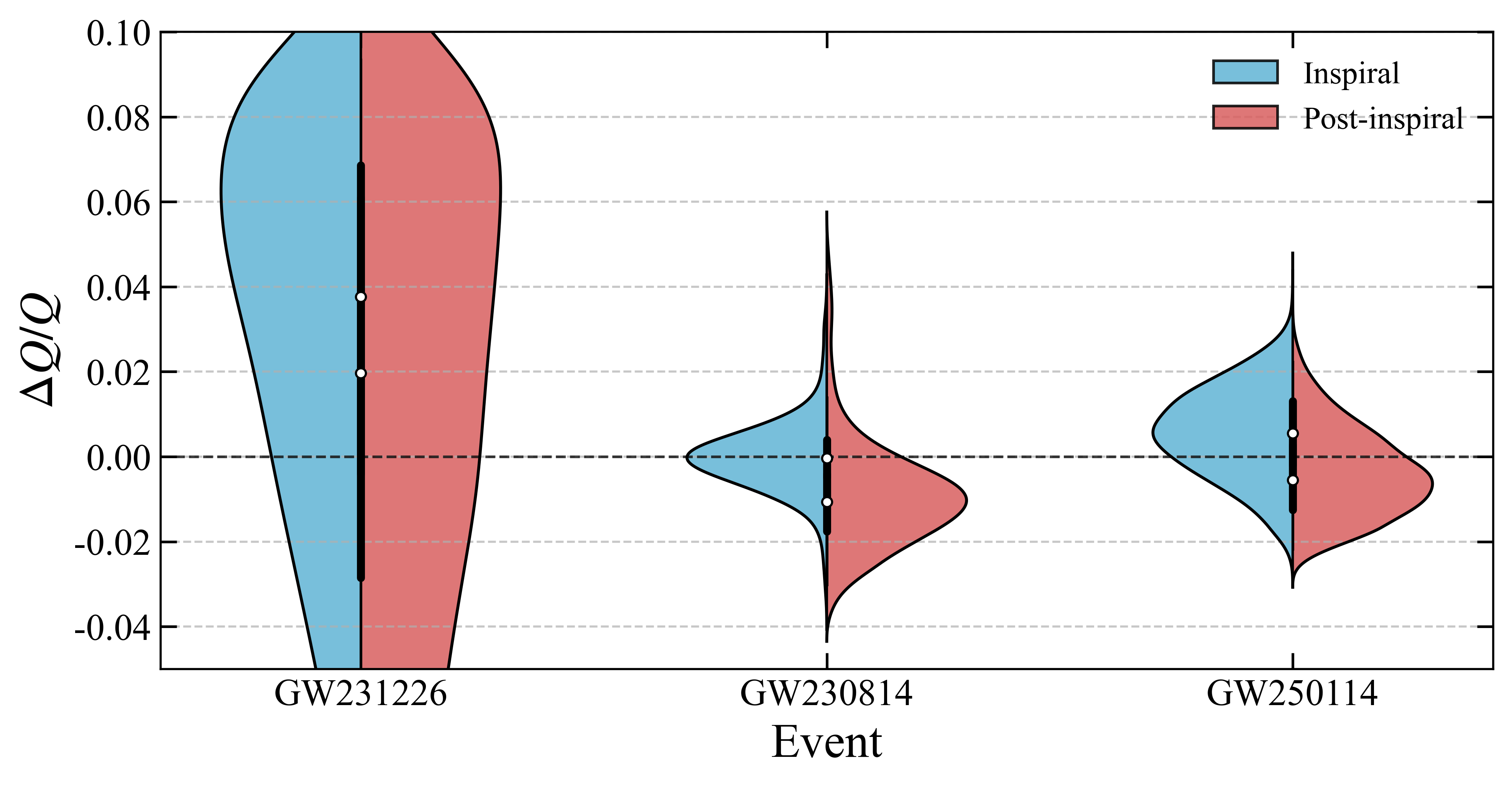}
\caption{Posterior distributions of $\Delta Q/Q$ for different events obtained with the $\Psi_{\mathrm{FD}}$ model using inspiral (blue) and post-inspiral (red) signals. The violin plots show kernel density estimates of the posterior samples. White dots indicate the medians, and the thick and thin vertical lines represent the 50\% and 90\% credible intervals. The dashed line at $\Delta Q/Q=0$ denotes the prediction of general relativity.}\label{DeltaQ_Qrea_inspiral_postinspiral}
\end{figure*}

Fig.~\ref{DeltaQ_Qrea_inspiral_postinspiral} presents the posterior distributions of the relative quadrupole moment deviation $\Delta Q/Q$ obtained from separate inspiral and post-inspiral analyses of the selected events. The blue distributions correspond to the inspiral analyses, while the red distributions correspond to the post-inspiral analyses.

For GW231226, $\Delta Q/Q$ is only weakly constrained in both frequency regimes, which may be partly attributed to the comparatively low SNRs of the inspiral and post-inspiral portions of the signal. By contrast, for GW230814 and GW250114, both the inspiral and post-inspiral analyses yield posterior distributions centered close to $\Delta Q/Q=0$.

In the inspiral analysis, the posterior for GW230814 is centered closer to zero than that for GW250114, despite its lower inspiral SNR. This difference may be related to their effective spins. In particular, GW230814 has a smaller effective-spin magnitude, $|\chi_{\mathrm{eff}}|\simeq 0.02$, compared with $|\chi_{\mathrm{eff}}|\simeq 0.07$ for GW250114. As discussed above, lower spin magnitudes reduce the spin-induced quadrupolar imprint in the inspiral waveform, thereby weakening the sensitivity to quadrupole moment deviations. The near-zero result for GW230814 should therefore not be interpreted as providing stronger evidence for the no-hair theorem.

In the post-inspiral analysis, a different trend is observed: the posterior for GW250114 is centered closer to zero than that for GW230814. Since the remnant spins of the two events are comparable, this difference may be more closely associated with the relative strength of the post-inspiral signals. In particular, the higher post-inspiral SNR of GW250114 may lead to a more informative constraint on $\Delta Q/Q$ in this regime.

Taken together, the inspiral and post-inspiral results for GW230814 and GW250114 are both compatible with $\Delta Q/Q=0$ and therefore show no observable departure from the no-hair theorem within the current measurement precision. Although the interpretation remains limited by the small number of events, these results demonstrate the potential of high-SNR observations to constrain deviations in the quadrupole moment. Future detections of additional binary black hole mergers with high SNRs, particularly those with sufficiently large spins, will be essential for obtaining more precise and robust tests of the no-hair theorem.

\section{Conclusion}\label{Conculsion_sec}
In this work, we tested the black hole no-hair theorem using five binary black hole merger events observed by the LIGO-Virgo-KAGRA Collaboration. We employed the $\Psi_{\mathrm{FD}}$ model, a beyond-general-relativity full-waveform template that introduces a parametrized deviation of the quadrupole moment from the Kerr prediction. Extending our previous analysis of GW150914 and GW200129, we considered three additional events, GW231226, GW230814, and GW250114, and compared the inferred source parameters with those obtained using the $\texttt{IMRPhenomXPHM}$ waveform model. The two waveform models yield broadly consistent constraints on the component masses, spins, and luminosity distances, with the remaining differences possibly arising from the additional quadrupole-deviation parameter included in $\Psi_{\mathrm{FD}}$.

For GW200129, GW150914, GW231226, GW250114, and GW230814, we obtained relative quadrupole moment deviations $0.07^{+0.02}_{-0.02}$, $0.06^{+0.03}_{-0.05}$, $0.04^{+0.02}_{-0.03}$, $0.00^{+0.01}_{-0.01}$, and $0.00^{+0.01}_{-0.01}$, respectively. The corresponding logarithmic Bayes factors comparing the $\Psi{\mathrm{FD}}$ and $\texttt{IMRPhenomXPHM}$ models are $5.10$, $1.59$, $0.62$, $-1.21$, and $0.04$. While GW200129 and GW150914 exhibit comparatively larger nonzero deviations, the three more recent events provide less pronounced indications of a departure from the Kerr quadrupole moment. In particular, the results for GW230814 and GW250114 are consistent with $\Delta Q/Q=0$ within the current measurement precision. 

Our comparison of the inferred quadrupole moment deviations and event SNRs suggests that high-SNR observations generally provide more informative constraints on $\Delta Q/Q$. Nevertheless, the measurability of a quadrupole moment deviation is not controlled by the signal strength alone. The source spin also plays an important role because the spin-induced quadrupolar imprint becomes weaker in low-spin systems. This effect may contribute to the near-zero deviation inferred for GW230814, whose effective spin is close to zero. Consequently, its result cannot be interpreted as either a particularly stringent confirmation of the Kerr prediction or solely as a consequence of reduced sensitivity to quadrupole moment deviations. GW250114 provides a complementary case: despite having the highest SNR in the sample and a larger effective-spin magnitude, its inferred deviation also remains close to zero. This result is compatible with the Kerr prediction, although a definitive interpretation requires a larger sample of high-SNR events.

We further investigated these constraints by performing separate parameter-estimation analyses of the inspiral and post-inspiral portions of the signals. For GW231226, $\Delta Q/Q$ remains only weakly constrained in both frequency regimes, which may be related to the comparatively modest inspiral and post-inspiral SNRs. By contrast, the inspiral and post-inspiral results for both GW230814 and GW250114 are centered close to zero and show no clear inconsistency between the two stages of the coalescence. In the inspiral regime, the result for GW230814 may be influenced by its smaller effective spin and correspondingly weaker quadrupolar imprint. In the post-inspiral regime, the higher signal strength of GW250114 leads to a more informative constraint than that obtained for GW230814.

Overall, the full-waveform, inspiral, and post-inspiral analyses of GW230814 and GW250114 reveal no observable departure from the no-hair theorem within the sensitivity of the current data and the $\Psi_{\mathrm{FD}}$ framework. However, the limited number of events prevents a definitive conclusion. As the gravitational-wave catalog continues to grow in future observing runs, additional high-SNR binary black hole mergers are expected to be detected. Such events will enable increasingly stringent and robust tests of the black hole no-hair theorem.

\begin{acknowledgments}
This work is supported by The National Science and Technology Major Project of China (No. 2024ZD1100601), The National Key R\&D Program of China (Grant No. 2021YFC2203002), NSFC (National Natural Science Foundation of China) No. 12473075, No. 12173071. This work made use of the High Performance Computing Resource in the Core Facility for Advanced Research Computing at Shanghai Astronomical Observatory.
\end{acknowledgments}

\bibliographystyle{apsrev4-1} 

\end{document}